\documentstyle[12pt]{article}
\begin{document}

\title{The Sliding-singlet Mechanism Revived\footnote{Supported 
in part by Department of Energy grant
\#DE-FG02-91ER406267}}

\author{{\bf S.M. Barr}\\Bartol Research Institute\\
University of Delaware\\Newark, DE 19716}

\date{BA-97-17}
\maketitle

\begin{abstract}

It is shown, using a modification of an idea of
Sen, that completely realistic supersymmetric
grand-unified theories based on $SU(6)$ or larger
unitary groups can be constructed using the
sliding-singlet mechanism. These models have a simple
structure, preserve the successful prediction of
$\sin^2 \theta_W$, and can suppress Higgsino-mediated
proton decay to an acceptable level in a simple way.

\end{abstract}

\section{Introduction}

The impressive unification of gauge couplings$^{1}$ at a scale
of $10^{16}$ GeV in the supersymmetric standard model
has led to renewed interest in the idea of supersymmetric
grand unification. The main theoretical difficulty with
grand unified theories has always been the gauge hierarchy
problem,$^{2}$ of which a key aspect is the so-called ``doublet-triplet
splitting problem".$^{3}$ This refers to the fact that in grand-unified
theories the color-triplet scalar that is in a unified multiplet
with the Higgs doublet of the Standard Model must be superheavy
to avoid rapid proton decay, while the Higgs doublet itself must have
a mass near the Weak-interaction scale.

Four interesting and elegant ways to 
achieve natural doublet-triplet splitting have been proposed.
These are the ``sliding-singlet mechanism"$^{4}$, the ``missing 
partner mechanism"$^{5}$, the ``Dimopoulos-Wilczek mechanism"$^{6}$ 
(also called the ``missing-vacuum-expectation-value mechanism"),
and the ``GIFT mechanism"$^{7}$. Each of these
ideas has notable strengths and weaknesses. The-missing partner
mechanism is the only one which works in $SU(5)$, the smallest
unified gauge group, but requires in $SU(5)$ the existence 
of Higgs fields in the
high-rank tensor representations ${\bf 50}$, $\overline{{\bf 50}}$,
and ${\bf 75}$.$^{5}$ The same mechanism works very elegantly
in the flipped $SU(5) \times U(1)$ group,$^{8}$ but as this group is
not fully unified the sharp prediction of gauge-coupling unification
is lost. 

The Dimopoulos-Wilczek mechanism is the only one which works in 
$SO(10)$, regarded by many as the most attractive candidate for
the grand-unified group. However, for such models to be fully
realistic it seems that the Higgs sector must be somewhat 
involved.$^{9}$ The ``GIFT" mechanism (in which the Higgs doublet
is light because it is a pseudo-goldstone field) solves the 
doublet-triplet problem in a very simple way in the group $SU(6)$,
but has the disadvantage that the quarks and leptons must get mass in 
a somewhat complicated fashion.$^{10}$

The first idea mentioned, the sliding-singlet mechanism, is perhaps the
prettiest of all, but was shown to have a serious difficulty that
prevents it from working in $SU(5)$. In particular, the gauge hierarchy
is destroyed by radiative corrections after supersymmetry breaks.$^{11}$
Later A. Sen$^{12}$ showed that the sliding-singlet mechanism can work 
in the group SU(6) to give a stable hierarchy. A shortcoming of his
model, however, in the light of later, 
more precise measurements of the gauge
couplings is that it introduces an intermediate scale
into the sequence of gauge-group breaking. $SU(6)$ breaks 
to $SU(3) \times SU(3) \times U(1)$ at a scale of order $10^{17}$ GeV, 
and then to $SU(3)_c \times SU(2)_L \times U(1)_Y$ at an intermediate
scale of order $10^{10}$ GeV. This gives a value of $\sin^2 \theta_W$
which is $0.211 \pm 0.003$, or seven standard deviations from the
presently measured value of $0.2324 \pm 0.0003$, and in fact slightly
worse than the minimal {\it non-supersymmetric} $SU(5)$ model. 

In this letter we show that a simple twist on Sen's idea allows a
fully realistic implementation of the sliding singlet mechanism
in $SU(6)$ and larger unitary groups. Before describing this improvement
we will briefly explain the main ideas in the previous development 
of the sliding-singlet mechanism.

\section{The Sliding-singlet Mechanism}

\noindent
{\bf (1) The basic idea.}

The basic idea of the sliding singlet mechanism as first proposed$^4$
in $SU(5)$ is based on the existence of terms of the following kind
in the Higgs superpotential:

\begin{equation}
W_{2/3} = \overline{H} \cdot ( \Sigma + S) \cdot H.
\end{equation}

\noindent
Here, $\Sigma$ is an adjoint Higgs field (${\bf 24}$), $\overline{H}$
and $H$ are an anti-fundamental and fundamental ($\overline{{\bf 5}}
+ {\bf 5})$, and $S$ is a singlet. It is assumed that some other
set of terms, $W(\Sigma)$, in the superpotential (there are many 
possibilities for these terms) gives $\langle \Sigma \rangle = 
{\rm diag}( -\frac{2}{3}
\Sigma_0, -\frac{2}{3} \Sigma_0, -\frac{2}{3} \Sigma_0, \Sigma_0, 
\Sigma_0)$, which breaks $SU(5)$ down to $SU(3) \times SU(2) \times
U(1)$. Then the equation, $F_{\overline{H}} = \partial W/\partial
\overline{H} = 0$, which is valid at a supersymmetric minimum, gives
$(\langle \Sigma \rangle - \langle S \rangle) \cdot \langle H \rangle 
= 0$. Since the $SU(2)$-doublets in $\overline{H}$ and $H$ are supposed to 
do the $SU(2) \times U(1)$ breaking of the Standard Model, they have
vacuum expectation values which are non-vanishing:
$v_1$ and $v_2$, where $\left| v_1 
\right|^2 + \left| v_2 \right|^2 \equiv v^2 = (246 {\rm GeV})^2$. 
This means
that the $F_{\overline{H}} = 0$ equation implies that $\langle S 
\rangle = - \Sigma_0$ and therefore 

\begin{equation}
\langle \Sigma \rangle + \langle S \rangle = {\rm diag}(-\frac{5}{3}
\Sigma_0, -\frac{5}{3} \Sigma_0, -\frac{5}{3} \Sigma_0, 0, 0).
\end{equation}

\noindent
The $F_H = 0$ equation gives the same result. The mechanism receives
its name from the fact that the singlet slides to cancel off the 
expectation value of the adjoint in the $SU(2)$ block. As a result
of the form in Eq. (2), the terms in Eq. (1) give superheavy ($\Sigma_0
\approx 10^{16}$ GeV) mass to the color triplets in $\overline{H}$ and 
$H$, while leaving the doublets massless --- the desired $2/3$ 
splitting.

\vspace{1cm}

\noindent
{\bf (2) The stability of the hierarchy}

This mechanism breaks down when account is taken of the breaking of
supersymmetry.$^{11}$ The potential that makes the singlet ``slide" is 
weak: the terms $\left| F_{\overline{H}} \right|^2 + \left| F_H
\right|^2$ give only a mass of order $v^2$ to
$S$. The supersymmetry-breaking contributions
to the potential of $S$ are of the same order as this and therefore 
disrupt the cancellation between $\Sigma$ and $S$. More specifically,
because $S$ couples to the superheavy triplets in $\overline{H}$ and
$H$, one-loop tadpole graphs$^{11}$ which have these
triplets running around the loop induce in the low-energy effective 
theory two terms that destroy the gauge hierarchy. These are
$T_1 = O(m_g^2 M_G) S + H.c.$ and $T_2 = O(m_g M_G) F_S + H.c.$, 
where $M_G$ is the unification scale, and $m_g$ the gravitino mass,
which is of order the Weak scale.

The term $T_1$, when added to the supersymmetric piece of the 
potential for $S$,
$(\left| \langle \overline{H} \rangle \right|^2 + \left| \langle H 
\rangle \right|^2) \left| S + \Sigma_0 \right|^2$, will evidently
shift the expectation value of $S$ from its supersymmetric minimum
at $- \Sigma_0$ by an amount of order $m_g^2 M_G/v^2 \sim M_G$, and thus
the term in Eq. (1) will give the doublets 
in $\overline{H}$ and $H$ superheavy mass. Moreover, 
the term $T_2$ will (after eliminating the auxiliary field $F_S$)
give a potential to the doublets of the
form $\left| \overline{H} H + O(m_g M_G) \right|^2$, which 
also is evidently incompatible with the gauge hierarchy.

\vspace{1cm}

\noindent
{\bf (3) The SU(6) Model of Sen}

In 1984 A. Sen made the clever observation$^{12}$ that the 
sliding-singlet
mechanism can be made stable to supersymmetry-breaking
radiative effects in groups of larger rank, like $SU(6)$. The essential
point is that the expectation values of $\overline{H}$ and
$H$ that force the singlet to slide can now be those which break
$SU(6)$ down to $SU(5)$, which are very large compared to the 
supersymmetry-breaking scale,
rather than those which break $SU(2)_L \times U(1)_Y$. 
Thus the supersymmetric part of the potential for the sliding singlet
is made more rigid and less subject to disruption by 
supersymmetry-breaking effects.

The relevant terms have the same form as in Eq. (1), with now, of 
course the adjoint $\Sigma$ being a ${\bf 35}$ and the fundamentals
$\overline{H}$ and $H$ being $\overline{{\bf 6}} + {\bf 6}$. 

Suppose that some terms in the superpotential  
cause $\langle \Sigma \rangle$ to have the form
$\langle \Sigma \rangle = {\rm diag}( - \Sigma_0, - \Sigma_0, - \Sigma_0,
+ \Sigma_0, + \Sigma_0, + \Sigma_0)$. Let the VEVs of the
standard-model-singlet components of $\overline{H}$ and $H$ be
$\langle \overline{H}_6 \rangle = \langle H^6 \rangle = H_0 \gg M_W$.
Then, in the supersymmetric limit, the equations $F_H = 0$ 
and $F_{\overline{H}} = 0$ yield the
same condition $\langle S \rangle = - \Sigma_0$ as before.

With supersymmetry breaking, the addition of the term $T_1$ to the potential
for $S$ gives
$V(S) = 2 H_0^2 \left| S + \Sigma_0 \right|^2 + (O(m_g^2
M_G) S + H.c.)$. This leads to a shift of the VEV of $S$ from the value 
$- \Sigma_0$ by an amount of order
$m_g^2 M_G/H_0^2$. So that this may be no larger than the
Weak scale it is only necessary that $H_0 \stackrel{_>}{_\sim}
\sqrt{m_g M_G} \sim 10^9$ GeV. Similarly, the term $T_2$
will lead to a contribution to the potential of the form
$\left| \overline{H} H + O(m_g M_G) \right|^2$, as already noted above.
This by itself would induce a
VEV for $H$ and $\overline{H}$ of order $\sqrt{m_g M_G}$. But since the
`6' components of these fields are assumed to have VEVs this large
anyway, the term $T_2$ poses no problem for the gauge hierarchy.

There are, however, three apparent problems that the sliding singlet
mechanism faces in $SU(6)$, and it is instructive to see how they 
are resolved in Sen's scheme. (a) From Eq. (1) it appears that the 
$F_S = 0$ equation 
forces $\langle \overline{H} \rangle = \langle H \rangle = 0$. 
(b) While the form of $\langle (\Sigma + S) \rangle$ means that it
does not contribute to the masses of the $SU(2)$ doublets, the VEVs
$\langle H^6 \rangle$ and $\langle \overline{H}_6 \rangle$ do. In
particular, one has from Eq. (1): $\overline{H}_i \Sigma^i_6 
\langle H^6 \rangle$
and $ \langle \overline{H}_6 \rangle \Sigma^6_i H^i$, where $i = 1,2$
are the SU(2) indices. (c) The term of Eq.(1) makes a contribution
to $F_{\Sigma} \equiv \partial W/ \partial \Sigma$ of $\langle 
H \overline{H} \rangle = {\rm diag}(0,0,0,0,0, H_0^2)$. This creates the
danger that the form of the VEV of $\langle \Sigma \rangle$ necessary
for the sliding-singlet mechanism to work (specifically, that
$\langle \Sigma^i_i \rangle  = \langle \Sigma^6_6 \rangle$)
would be destabilized.

As for (a), in Sen's model, the $F_S = 0$ equation does indeed imply that 
$\langle \overline{H} \rangle = \langle H \rangle = 0$ is the
correct vacuum in the supersymmetric limit. But, as already mentioned,
including supersymmetry-breaking effects gives 
just the term $\left| \overline{H} H + O(m_g M_G) \right|^2$,
and leads to $H_0 \sim \sqrt{m_g M}$. As we have seen
this is large enough for the hierarchy not to be destabilized.

As for (b), in Sen's model, there are indeed mass terms of order $H_0$ 
connecting the doublets in the fundamentals with the doublets in the
adjoint. But there are also mass terms of order $\Sigma_0 \sim M_G$
connecting the doublets in the adjoints to themselves. Thus by a
``see-saw mechanism", there are doublets which are eigenstates with mass
of order $H_0^2/\Sigma_0 \sim m_g \sim M_W$. 
For details, readers are referred to
Ref. (12).

Finally, as for (c), in Sen's model the contribution of $\langle H 
\overline{H} \rangle$ to $F_{\Sigma}$ is only of order $m_g M_G$, and
so the form of $\langle  \Sigma \rangle$ required for the sliding
singlet mechanism to work is only shifted by $O(m_g)$, preserving
the hierarchy.

From the foregoing, it is clear that 
the vacuum expectation values of the $H$ and $\overline{H}$
being at the intermediate scale $m_g M_G \sim 10^9$ GeV rather
than at the GUT scale is crucial in the model of Sen.
What this means is that at the grand unification scale $SU(6)$
breaks to $SU(3) \times SU(3) \times U(1)$, which then breaks
to the standard model group at $10^{9}$ GeV. The consequence is
that $\sin^2 \theta_W$ is predicted to be $0.211 \pm 0.003$, which
as noted in the Introduction is far from the presently observed value.

\section{A Satisfactory Sliding-singlet Mechanism}  

We shall now describe an implementation of the sliding-singlet
mechanism in $SU(6)$, which incorporates the
essential idea of Sen, but in which the unified group breaks
all the way to the Standard Model at the unification scale of
$10^{16}$ GeV. This idea can be generalized to all $SU(N)$, 
with $N \geq 6$.

Let the Higgs superpotential of an $SU(6)$ model have the form

\begin{equation}
\begin{array}{ccl}
W & = & W( \Sigma) + W(\overline{H}_A, H_A) \\
& & \\
& & + \sum_{A = 1,2} \lambda_A \overline{H}_A( \Sigma + S_A) h_A
+ \sum_{A = 1,2} \overline{\lambda}_A \overline{h}_A 
(\Sigma + \overline{S}_A) H_A.
\end{array}
\end{equation}

\noindent
$W(\Sigma)$ is some set of terms that has as one of its discrete
set of minima $\langle \Sigma \rangle = {\rm diag}
(- \Sigma_0, - \Sigma_0, - \Sigma_0, + \Sigma_0, + \Sigma_0,
+ \Sigma_0)$. One possibility is $W(\Sigma) = \Sigma^3 + M \Sigma^2
+ X(\Sigma^2 - m^2)$, where $X$ is a singlet.
Another possibility, which is simpler but uses a higher-dimensional
operator is $W(\Sigma)= \Sigma^4 - M^2 \Sigma^2$.
$W(\overline{H}_A, H_A)$ is such as to give the fields 
$\overline{H}_A$ and $H_A$ vacuum expectation values of order $M_G$
pointing in the `6' direction. These fundamentals together
with the adjoint break $SU(6)$ all the way down to $SU(3)_c \times
SU(2)_L \times U(1)_Y$ at the scale $M_G \sim 10^{16}$ GeV.

The equations $F_{\overline{h}_A} = 0$ and $F_{h_A} = 0$
force the singlets $S_A$  and $\overline{S}_A$ to slide so that 
$\langle S_A \rangle = \langle \overline{S}_A \rangle =
- \Sigma_0$, as before. 

A critical point is that the fields $\overline{h}_A$ and
$h_A$, which are in the representations $\overline{{\bf 6}} + {\bf 6}$,
have vanishing expectation values in the supersymmetric limit.
(As we shall see, they will get expectation values of order $m_g$
when supersymmetry breaks.) This allows a simple resolution of the
three potential difficulties mentioned in the last section.

(a) The equation $F_{S_A} = 0$ implies that $\langle \overline{H}_A
h_A \rangle = 0$ in the supersymmetric limit. But in contrast to
Sen's model, this is here not at all inconsistent with $\overline{H}_A$
having an expectation value of order $M_G$, since it can be satisfied
by $\langle h_A \rangle = 0$. When supersymmetry-breaking effects
are included one has $V(S) = \left| \overline{H}_A h_A + O(m_g
M_G) \right|^2$. This merely induces an expectation value of order
$m_g$ in the `6' component of $h_A$. (It also induces a mass-squared
term of order $m_g M_G$ connecting the scalar doublets in $\overline{H}$ 
and $h$, but since $\overline{H}$ has a mass of order $M_G$,
as will be seen shortly, this only gives a contribution of order 
$m_g^2$ to the light doublet, which turns out to be in $h$.) 
The same discussion applies,
of course to the $F_{\overline{H}_A} = 0$ equation.
 
The foregoing also resolves potential difficulty (c). The contribution
of the last two terms in Eq. (3) to $F_{\Sigma}$ is 
$\sum_A \lambda_A h_A \overline{H}_A  + \sum_A 
\overline{\lambda}_A H_A \overline{h}_A$.
But, as in Sen's model, this is of order $m_g M_G$. The result
is that the vacuum expectation value of $\Sigma$ is shifted by
order $m_g$, leaving the gauge hierarchy intact.

Potential difficulty (b) was that the expectation values
of $\overline{H}_A$ and $H_A$ would give superlarge mass terms
that connect doublets in the adjoint $\Sigma$ to those in the
fields $h_A$ and $\overline{h}_A$. However, as there is only
one doublet with quantum numbers $(1,2, -\frac{1}{2})$ in the adjoint,
only one linear combination of the two doublets with $(1,2,
\frac{1}{2})$ that are contained in $h_1$ and $h_2$ is made
superheavy, the orthogonal linear combination being the light
Higgs multiplet; and similarly for the conjugate doublets.
The situation is made clear by examining the full doublet mass
matrix.

\begin{equation}
\begin{array}{l}
W_{{\rm doublet mass}} = 
\left( \Sigma_i^6, (\overline{h}_1)_i, (\overline{h}_2)_i,
(\overline{H}_1)_i, (\overline{H}_2)_i \right) \cdot \\
\\
\left( \begin{array}{ccccc}
M_{\Sigma} & \lambda_1 \langle \overline{H}_1 \rangle &
\lambda_2 \langle \overline{H}_2 \rangle & 0 & 0 \\
\overline{\lambda}_1 \langle H_1 \rangle & 0 & 0 & 0 & 0 \\
\overline{\lambda}_2 \langle H_2 \rangle & 0 & 0 & 0 & 0 \\
0 & 0 & 0 & c \left| \langle H_2 \rangle
\right|^2 & -c \langle H_1 H_2^* \rangle \\
0 & 0 & 0 & -c \langle H_1^* H_2 \rangle & c \left| \langle H_1 \rangle 
\right|^2  
\end{array} \right) \left( 
\begin{array}{c}
\Sigma_6^i \\ (h_1)^i \\ (h_2)^i \\ (H_1)^i \\ (H_2)^i
\end{array} \right).
\end{array}
\end{equation}

\noindent
Here `$i$' is an $SU(2)_L$ index. 
The parameters $M_{\Sigma}$ and $c$ depend on the details of 
$W(\Sigma)$ and $W(\overline{H}_A, H_A)$ respectively. $c$ has
dimensions of inverse mass, and typically the mass of the $H$ and
$\overline{H}$ fields goes as $c \left| \langle H \rangle \right|^2$.
One sees from the form of the matrix
that the goldstone doublets that are eaten in the breaking of
$SU(6)$ down to the Standard Model are contained in the 
$\overline{H}_A$ and $H_A$, while the light doublets that are the
Higgs of the Standard model are contained in the $\overline{h}_A$
and $h_A$. It should be noted that due to the shifts caused by
supersymmetry-breaking, some of the zeros in Eq. (4) 
are really non-vanishing and of order $m_g$.

The mass matrix of the color-triplet Higgs is similar in form.

\begin{equation}
\begin{array}{l}
W_{{\rm doublet mass}} = 
\left( \Sigma_a^6, (\overline{h}_1)_a, (\overline{h}_2)_a,
(\overline{H}_1)_a, (\overline{H}_2)_a \right) \cdot \\
\\
\left( \begin{array}{ccccc}
0 & \lambda_1 \langle \overline{H}_1 \rangle &
\lambda_2 \langle \overline{H}_2 \rangle & 0 & 0 \\
\overline{\lambda}_1 \langle H_1 \rangle & 0 & 0 & 
-2 \overline{\lambda}_1 \Sigma_0 & 0 \\
\overline{\lambda}_2 \langle H_2 \rangle & 0 & 0 & 0 & 
-2 \overline{\lambda}_2 
\Sigma_0 \\
0 & -2 \lambda_1 \Sigma_0 & 0 & c \left| \langle H_2 \rangle
\right|^2 
& -c \langle H_1 H_2^* \rangle \\
0 & 0 & -2 \lambda_2 \Sigma_0 & -c \langle H_1^* H_2 \rangle 
& c \left| \langle H_1 \rangle \right|^2 
\end{array} \right) \left( 
\begin{array}{c}
\Sigma_6^a \\ (h_1)^a \\ (h_2)^a \\ (H_1)^a \\ (H_2)^a
\end{array} \right).
\end{array}
\end{equation}

Here `$a$' is a color index. There is only one zero-eigenvalue of this
matrix corresponding to the goldstone mode that is eaten in the breaking
of $SU(6)$ down to the Standard Model, namely $(2 \Sigma_0, 0, 0,
\langle H_1 \rangle, \langle H_2 \rangle)$. 
Thus natural doublet-triplet splitting has been achieved. 

It is interesting to see how the amplitude for Higgsino-mediated
proton decay, which is generally a problem for supersymmetric
grand unified theories, depends on the parameters of the model.
From the matrices given in Eqs. (4) and (5) it is straighforward to
derive that the propagator of the colored Higgsino that mediates 
proton decay is given by

\begin{equation}
(M_3)^{-1} = -\frac{c}{4 \Sigma_0^2} \left[
\frac{(\sqrt{\lambda_2 \overline{\lambda}_2/\lambda_1 
\overline{\lambda}_1} \left| \langle H_1 \rangle \right|^2 
- \sqrt{\lambda_1 \overline{\lambda}_1/\lambda_2 
\overline{\lambda}_2} \left| \langle H_2 \rangle \right|^2)^2}
{(\left| \overline{\lambda}_1 \langle H_1 \rangle \right|^2 
+ \left| \overline{\lambda}_2 \langle H_2 \rangle 
\right|^2)^{\frac{1}{2}}
(\left| \lambda_1 \langle H_1 \rangle \right|^2 
+ \left| \lambda_2 \langle H_2 \rangle 
\right|^2)^{\frac{1}{2}}} \right].
\end{equation}

This is to be compared to the value $(M_3)^{-1} \sim 
(\Sigma_0)^{-1}$ that one gets in the (fine-tuned) minimal
supersymmetric $SU(5)$ model. One sees that there is an extra factor
in the Higgsino-mediated proton-decay amplitude which is
of order $M_H/\Sigma_0$, where we recall that $M_H \sim c \left| \langle
H \rangle \right|^2$. If this factor is of order $10^{-1}$
then the Higgsino-mediated proton-decay
rate is comfortably suppressed below present bounds.

In this model there are, altogether, in the Higgs sector 
a ${\bf 35} + 4 ( {\bf 6} + \overline{{\bf 6}})$. (This is 
compared to a ${\bf 35} + 2 ( {\bf 6} + \overline{{\bf 6}})$ in
Sen's model.) In terms of multiplets of the $SU(5)$ subgroup
there are ${\bf 24} + 5 ( {\bf 5} + \overline{{\bf 5}})$ as well as some
singlets. One pair of ${\bf 5} + \overline{{\bf 5}}$ gets eaten by
the gauge bosons in $SU(6)/SU(5) \times U(1)$. Thus, the model
differs from minimal supersymmetric $SU(5)$, as far as computing 
$\sin^2 \theta_W$ is concerned, by the presence of three additional
scalar multiplets and one additional gauge multiplet of $({\bf 5} 
+ \overline{{\bf 5}})$. All of the components of these extra multiplets
are superheavy, and as they are small representations, they
have only a minor effect on $\sin^2 \theta_W$. One can show that
the shift from the prediction of (fine-tuned) minimal supersymmetric
$SU(5)$ is $\Delta \sin^2 \theta_W \cong \frac{3 \alpha(M_Z)}{10 \pi}  
(5 \ln (\Sigma_0/\langle H \rangle) + \ln (\Sigma_0/M_H))$. 

If the expectation values of the adjoint and fundamental Higgs fields
that break $SU(6)$ are within a factor of three of each other, then
the first term in the parentheses gives a typical threshold 
correction of about $\pm 0.005$. The second term in parentheses is
interesting since the argument of the logarithm is essentially 
the suppression factor of the Higgsino-mediated-proton-decay
amplitude. Thus a suppresion of the proton decay rate by 
factor of $10^{-2}$ below
the minimal supersymmetric $SU(5)$ level would give a shift of 
$\sin^2 \theta_W$ upward by about $0.002$, which is negligible.

The quark and lepton masses can arise in a straightforward way.
The down-type quarks and charged leptons can get mass from
a ${\bf 15} \; \overline{{\bf 6}} \; \overline{{\bf 6}}$ term:
$\psi^{\alpha \beta} \psi_{\alpha} \overline{h}_{\beta}$, where we
have suppressed flavor indices. This is just the analogue of the
${\bf 10} \; \overline{{\bf 5}} \; \overline{{\bf 5}}$ term in 
minimal $SU(5)$. The up-type quarks (if there is minimal quark and
lepton content) get mass from the dimension-5 operator
$\psi^{\alpha \beta} \psi^{\gamma \delta} h^{\zeta} H^{\eta}
\epsilon_{\alpha \beta \gamma \delta \zeta \eta}/M_G$, where again we
have suppressed flavor indices. When the $H$ gets an expectation value 
of order $M_G$ in the `6' direction, this term reduces to the
ordinary ${\bf 10} \; {\bf 10} \; {\bf 5}$ coupling of minimal $SU(5)$.
In other words, the quarks and leptons get mass as in a ``minimal
$SU(6)$ model". The sliding-singlet mechanism in no way complicates
the issue of light fermion masses as it does in the ``GIFT" approach.

It can be shown that the gauge hierarchy can be made
stable to the effect of higher-dimension operators in the Higgs sector. 
There are two kinds of operators that must be excluded from the
Higgs superpotential. These are $\overline{h}_A h_B$ terms
that would directly give mass to the light doublets, and 
$\overline{H}_A ( \Sigma \; {\rm or} \; S) H_B$ terms, which would 
create the difficulties (a)--(c) discussed in the last section.
It is straightforward to invent discrete or continuous symmetries
that forbid these kinds of terms to sufficiently high order
in $1/M_P$.

As a final comment on $SU(6)$ it might be asked whether one could
not modify Sen's model in a different way to make it realistic, by
simply adding another pair of fundamentals, $H' + \overline{H}'$, which
do not couple to the sliding singlets and which have a superpotential
that gives them VEVs of order $M_G$ that break $SU(6)$ to $SU(5)$, while
the fundamentals that participate in the sliding-singlet mechanism
get VEVs of order $\sqrt{m_g M_G}$. While it may be possible to construct
such models, they would face certain difficulties that would almost
certainly make them more complicated than the model we have
presented. In particular, if the fields $H'$ and $\overline{H}'$ do
not couple to the adjoint or the other fundamentals, there would be
unwanted goldstone bosons, while if they do they would tend to 
destabilize the VEVs of the adjoint or make the VEVs of the other
fundamental be of order $M_G$. 

The sliding-singlet mechanism in the realistic form described above
is immediately generalizable to any $SU(N)$ for $N > 6$. 

\section{Conclusions}

The sliding-singlet is perhaps the most elegant solution to the
doublet-triplet-splitting problem of grand-unified theories. 
We have shown that a perfectly realistic implementation of the
mechanism in $SU(6)$ and larger unitary groups can be achieved
by a variation on an old idea of A. Sen.

The sliding-singlet mechanism has certain advantages over other
approaches that have been proposed. The missing-partner mechanism
requires either large representations of Higgs to exist (in $SU(5)$)
or an abandonment of the precise and successful prediction of
$\sin^2 \theta_W$ (in flipped $SU(5) \times U(1)$). The ``GIFT"
mechanism makes it difficult to generate quark and lepton masses in a 
straighforward way. The Dimopoulos-Wilczek mechanism seems to require
(at least in $SO(10)$) a somewhat involved Higgs sector (though
it is the only mechanism that works in $SO(10)$, which may be the 
most promising group for grand unification from the point of
view of understanding the pattern of quark and lepton masses).

Looked at as a whole, grand unified models based on the sliding-singlet
mechanism as implemented here can claim to be the simplest in structure
that exist. The Higgs sector requires only a single adjoint and a set
of fundamentals and singlets. The prediction of $\sin^2 \theta_W$ is
undisturbed by large threshold corrections at the GUT scale, the
Higgsino-mediated proton-decay amplitude has automatically an extra
factor compared to minimal $SU(5)$ that allows it to be suppressed
to an acceptable level in a simple way. Both the Higgs sector and the
Yukawa sector are simple in structure. And the hierarchy can be
made stable to the effects of higher-dimension operators in
straightforward ways.

\vspace{1cm}

\noindent
{\bf Note Added:} After this work was completed the author became aware
of related work of G. Dvali, {\it Phys. Lett.}
{\bf 324B}, 59 (1994). Dvali's viewpoint is different and
involves the idea that Higgs doublets are light because they are related
to goldstone bosons by a custodial $SU(N)$ symmetry. His Higgs are
therefore in $({\bf 6}, {\bf N}) + H.c.$ of $SU(6) \times SU(N)$. 
He is led, however, to a structure similar to Eq. (3) 
of this paper. (See sec. 7 of Dvali's paper.)

\section*{References}

\begin{enumerate}
\item U. Amaldi, W. deBoer, and H. Furstenau, {\it Phys. Lett.}
{\bf 260B}, 447 (1991); P. Langacker and M.-X. Luo, {\it Phys. Rev.}
{\bf D44}, 817 (1991); J. Ellis, S. Kelley, and D.V. Nanopoulos,
{\it Phys. Lett.} {\bf 260B}, 131 (1991).
\item E. Gildener and S. Weinberg, {\it Phys. Rev.} {\bf D13}, 3333
(1976); E. Gildener, {\it Phys. Rev.} {\bf D14}, 1667 (1976).
\item L. Maiani, in {\it Comptes Rendus de l'Ecole d'Et\`{e} de Physiques
des Particules}, Gif-sur-Yvette, 1979, IN2P3, Paris, 1980, p3;
S. Dimopoulos and H. Georgi, {\it Nucl. Phys.} {\bf B150}, 193 (1981);
M. Sakai, {\it Z. Phys.} {\bf C11}, 153 (1981); E. Witten,
{\it Nucl. Phys.} {\bf B188}, 573 (1981).
\item E. Witten, {\it Phys. Lett.} {\bf 105B}, 267 (1981);
D.V. Nanopoulos and K. Tamvakis, {\it Phys. Lett.} {\bf 113B}, 151
(1982); S. Dimopoulos and H. Georgi, {\it Phys. Lett.} {\bf 117B},
287 (1982); L. Ibanez and G. Ross, {\it Phys. Lett.} {\bf 110B},
215 (1982)
\item H. Georgi, {\it Phys. Lett.} {\bf 108B}, 283 (1982);
A. Masiero, D.V. Nanopoulos, K. Tamvakis, and T. Yanagida, {\it Phys.
Lett.} {\bf 115B}, 380 (1982); B. Grinstein, {\it Nucl. Phys.}
{\bf B206}, 387 (1982).
\item S. Dimopoulos and F. Wilczek, Preprint NSF-ITP-82-07 (1982);
K.S. Babu and S.M. Barr, {\it Phys. Rev.} {\bf D48}, 5354 (1993).
\item K. Inoue, A. Kakuto, and T. Takano, {\it Prog. Theor. Phys.}
{\bf 75}, 664 (1986); A. Anselm and A. Johansen, {\it Phys. Lett.}
{\bf 200B}, 331 (1988); A. Anselm, {\it Sov. Phys. JETP} {\bf 67},
663 (1988); Z.G. Berezhiani and G. Dvali, {\it Sov. Phys. Lebedev
Inst. Reports} {\bf 5}, 55 (1989).
\item I. Antoniades, J. Ellis, J. Hagelin, and D.V. Nanopoulos,
{\it Phys. Lett.} {\bf 194B}, 231  (1987); {\it ibid.} {\bf 205B},
459 (1988).
\item K.S. Babu and S.M. Barr, Ref 6; {\it Phys. Rev.} {\bf D50},
3529 (1994); {\it Phys. Rev.} {\bf D51}, 2463 (1995).
\item R. Barbieri, G. Dvali, A. Strumia, Z. Berezhiani, and 
L.J. Hall, {\it Nucl. Phys.} {\bf B432}, 49 (1994);
Z.G. Berezhiani, {\it Phys. Lett.} {\bf 355B}, 481 (1995).
\item J. Polchinski and L. Susskind, {\it Phys. Rev.} {\bf D26},
3661 (1982); M. Dine, in {\it Lattice Gauge Theories, Supersymmetry,
and Grand Unification}, Proc. of the 6th Johns Hopkins Workshop,
Florence, Italy, 1982 (Johns Hopkins Press, Baltimore, 1982);
H.P. Nilles, M. Srednicki, and D. Wyler, {\it Phys. Lett.} 
{\bf 124B}, 237 (1982); A.B. Lahanas, {\it Phys. Lett.} {\bf 124B},
341 (1982).
\item A. Sen, {\it Phys. Lett.} {\bf 148B}, 65 (1984);
{\it Phys. Rev.} {\bf D31}, 900 (1985).
   
\end{enumerate}

\end{document}